\pdfoutput=1
\documentclass[journal]{IEEEtran}

\usepackage{graphicx}
\usepackage{psfrag}
\usepackage{epsfig}
\usepackage{ifthen}
\usepackage{law}
\usepackage{mystyle}
\usepackage{graphics}
\usepackage{times}
\usepackage{cite}

\usepackage{amsbsy} 
\usepackage{amsfonts} 
\usepackage{amsmath} 
\usepackage{amssymb} 
\usepackage{multicol}

\ifCLASSINFOpdf
\else
\fi

\begin{document}
%
\title{SVM and Dimensionality Reduction in Cognitive Radio with Experimental Validation }

\author{\IEEEauthorblockN{Shujie~Hou, 
Robert C. Qiu*, ~\IEEEmembership{Senior~Member,~IEEE,} Zhe Chen, Zhen Hu,~\IEEEmembership{Student~Member,~IEEE}
\thanks{The authors are with the Department of Electrical and Computer Engineering, Center
for Manufacturing Research, Tennessee Technological University, Cookeville, TN 38505, USA. E-mail: \{shou42, zchen42, zhu21
\}@students.tntech.edu, rqiu@tntech.edu.}}}
\maketitle



%
\IEEEpeerreviewmaketitle

\begin{abstract}

There is a trend of applying machine learning algorithms to cognitive radio. One fundamental open problem is to determine how and where these algorithms are useful in a cognitive radio network. In radar and sensing signal processing, the control of degrees of freedom (DOF)---or dimensionality---is the first step, called pre-processing. In this paper, the combination of dimensionality reduction with SVM is proposed apart from only applying SVM for classification in cognitive radio.  Measured Wi-Fi signals with high signal to noise ratio (SNR) are employed to the experiments. The DOF of Wi-Fi signals is extracted by dimensionality reduction techniques. Experimental results show that with dimensionality reduction, the performance of classification is much better with fewer features than that of without dimensionality reduction. The error rates of classification with only one feature of the proposed algorithm can match the error rates of 13 features of the original data.  The proposed method will be further tested in our cognitive radio network testbed.

\end{abstract}

\section{Introduction}

Intelligence and learning are  key factors to cognitive radio.  
Recently, there is a trend of applying machine learning algorithms to cognitive radio~\cite{clancy2007applications}. Machine learning~\cite{bishop2006pattern} is a discipline to design algorithms for computers to imitate human being's behaviors, which includes learning to recognize the complex patterns and making decisions based on experience automatically and intelligently. The topic of machine learning is precisely that needs to be introduced to cognitive radio. One fundamental open problem is to determine how and where these algorithms are useful in a cognitive radio network.  Such a systematical investigation is missing in the literature. It is the motivation of this paper to fill this gap. 

In radar and sensing signal processing, the control of degrees of freedom (DOF)---or intrinsic dimensionality---is the first step, called pre-processing. The network dimensionality, on the other hand, has received attention in information theory literature. One naturally wonders how network (signal) dimensionality affects the performance of system operation. Here we study, as an illustrative example, state classification of measured Wi-Fi signal in cognitive radio under this context.  Both linear such as principal component analysis (PCA) and nonlinear methods such as kernel principal component analysis (KPCA) and maximum variance unfolding (MVU) are studied, by combining them with  support vector machine (SVM)~\cite{vapnik2000nature,vapnik1998statistical,vapnik1997support,burges1998tutorial,smola2004tutorial,cristianini2000introduction}---the latest breakthrough in machine learning. Dimensionality reduction methods of PCA, KPCA and MVU  are systematically studied in this paper which can meet the needs of all kinds of  data owning different structures. The reduced dimension data can retain most of the useful information of the original data  but have much fewer dimensions.


SVM is both a linear and a nonlinear classifier which has been successfully applied in many areas such as handwritten digit recognition ~\cite{cortes1995support,schölkopf1995extracting,schölkopf1996incorporating} and object recognition~\cite{blanz1996comparison}. In cognitive radio, SVM has been exploited to do channel and modulation selection~\cite{xu2006channel}, signal classification~\cite{hu2008signal,ramón-signal,he2009signal} and spectrum estimation~\cite{atwood2009robust}. In~\cite{hu2008signal}, a method for combining feature extraction based on spectral correlation analysis with SVM to classify signals has been proposed. In this paper, SVM method will be explored as a classifier for measured Wi-Fi signal data.

Dimensionality reduction methods are innovative and important tools in machine learning~\cite{lee2007nonlinear}. The original dimensionality data collected in cognitive radio may contain a lot of features,  however, usually these features are highly correlated and redundant with noise. Thus the intrinsic dimensionality of the collected data is much fewer than the original features. Dimensionality reduction attempts to select or extract a lower dimensionality expression but retain most of the useful information. 

PCA~\cite{jolliffe2002principal} is the best-known linear dimensionality reduction method. 
PCA takes the variance among data as the useful information---it aims to find a subspace 
$\Omega $ 
which can maximally retain the variance of the original dataset. On the other hand, although linear PCA can work well (when such a subspace 
$\Omega $ 
exists), it always fails to detect the nonlinear structure of data. A nonlinear dimensionality reduction method called KPCA ~\cite{schölkopf1998nonlinear} can be used for this purpose.  KPCA uses the kernel tricks~\cite{scholkopf2002learning} to map the original data into a feature space ${\boldsymbol F}$, and then does PCA in ${\boldsymbol F}$ without knowing the mapping explicitly. The leading eigenvector of the sample covariance matrix in the feature space has been explored to spectrum sensing in cognitive radio ~\cite{hou2011spectrum}.

Manifold learning has become a very active topic in nonlinear dimensionality reduction.  The basic assumption for these manifold learning algorithms is that the input data lie on or close to a smooth manifold~\cite{seung2000manifold,murase1995visual}. The cognitive radio network happens to lie in this category.  A lot of promising methods have been proposed in this context,  including isomeric mapping (Isomap)~\cite{tenenbaum2000global}, local linear embedding (LLE)~\cite{roweis2000nonlinear}, Laplacian eigenmaps~\cite{belkin2003laplacian}, local tangent space alignment (LTSA)~\cite{zhang2004principal}, MVU ~\cite{weinberger2006unsupervised}, Hessian eigenmaps~\cite{donoho2003hessian}, manifold charting~\cite{brand2003charting}, diffusion maps~\cite{coifman2006diffusion} and Riemannian manifold learning (RML)~\cite{lin2006riemannian}. The MVU approach will be applied to our problem. MVU exploits semidefinite programming method---the latest breakthrough in convex optimization---to solve a convex optimization model that maximizes the output variance,  subject to the constrains of zero mean and local isometry.

As aforementioned, the collected data often contains too much redundant information. The redundant information not only complicates the algorithms but also conceals the underlying reason of the performance of the algorithms. Therefore, the intrinsic dimensionality  of the collected data should be extracted. In this paper, the combination of dimensionality reduction with SVM is also proposed apart from only applying SVM for classification. The intrinsic structure of the collected data in cognitive radio determines the uses of the corresponding linear or nonlinear dimensionality reduction methods.

The contributions of this paper are as follows. First, SVM is explored  to classify the states of Wi-Fi signal successfully. Second, the combination of SVM and dimensionality reduction is the first time proposed to cognitive radio, with the motivation stated above. Third, the measured Wi-Fi signal is employed to validate the proposed approach. In this paper, both the linear and nonlinear dimensionality reduction method will be systematically studied and applied to Wi-Fi signal.

We are building a cognitive radio network testbed~\cite{crn2010}. The dimensionality reduction techniques can be tested in the network testbed in real time. More applications, such as smart grid~\cite{crn2010,licommunication,licompressed}, and wireless tomography~\cite{tomo1,tomo2}, can benefit from the network testbed and machine learning techniques.

The organization of this paper is as follows. In section~\ref{svm}, SVM is briefly reviewed. Three different dimensionality reduction methods are introduced in section~\ref{dr}. 
The procedure of combining dimensionality reduction with SVM is revisited in section~\ref{ss}. Measure Wi-Fi signals are introduced in Section~\ref{wifi}. The experimental results are shown in section~\ref{ev}. Finally, the paper is concluded in Section~\ref{conc}.

\section {Support Vector Machine}
\label{svm}

Supervised learning is one of the three major learning types in machine learning. The dataset of supervised learning consists of $M$ pairs of inputs/outputs (labels) 
\begin{equation}
\label{inout}
({\bf x}_i ,l{}_i),i = 1,2, \cdots ,M.
\end{equation}
Suppose the training samples and testing samples are ${\bf x}_i$ with indexes $i_t, t=1,2,\cdots,T$ and $i_s,s=1,2,\cdots,S$, respectively. A supervised learning algorithm seeks mapping from inputs to outputs by training set and predicts any outputs based on the mapping. If the output is categorical or nominal value, then this will become a classification problem. Classification is a common task in machine learning.  SVM ~\cite{burges1998tutorial}~\cite{smola2004tutorial} methods will be employed in  two categories' classification experiments later.

Consider, for example, a simplest two classes classification task with the data that are linearly separable. Under this case,
\begin{equation}
l_{i_t } \in \left\{-1,1\right\}
\end{equation} indicates which class ${\bf x}_{i_t }$ belongs to. SVM attempts to find the separating hyperplane
\begin{equation}
\label{hyper}
{\bf w} \cdot {\bf x} + b = 0
\end{equation}
with the largest margin ~\cite{smola2004tutorial} satisfying the following constraints:
\begin{equation}
\label{constraint}
\begin{array}{l}
 {\bf w} \cdot {\bf x}_{i_t }  + b \ge 1{\rm  \:  \: \: \:\:for } \: \:l_{i_t }  = 1 \\ 
 {\bf w} \cdot {\bf x}_{i_t }  + b \le  - 1{\rm   \: \:  for }\: \:l_{i_t }  =  - 1 \\ 
\end{array}
\end{equation}
for linear separable case, in which $\bf w$ is the normal vector of the hyperplane and $\cdot$ stands for inner product. The constraints \eqref{constraint} can be combined into:
\begin{equation}
l_{i_t } ({\bf w} \cdot {\bf x}_{i_t }  + b) \ge 1{\rm  }.
\end{equation}

The requirements for separating hyperplane with the largest margin can formulate the problem into the following optimization model:
\begin{equation}
\label{op11}
\begin{array}{l}
 {\rm minimize  }\:\left\| {\bf w} \right\|^2  \\
 {\rm subject\: to }\\
   l_{i_t } ({\bf w} \cdot {\bf x}_{i_t }  + b) \ge 1 \\ 
 \end{array}
\end{equation}
in which $t=1,2,\cdots, T$. The dual form of \eqref{op11} by introducing Lagrange multipliers 
\begin{equation}
{\alpha} _{i_t} \geq 0,\:\:\:t=1,2,\cdots,T
\end{equation} 
is:
\begin{equation}
\begin{array}{l}
 {\rm maximize }\\
 \sum\limits_{i_t} {\alpha} _{i_t }  - \frac{1}{2}\sum\limits_{i_t,j_t} {\alpha} _{i_t} {\alpha}_{j_t} l_{i_t} l_{j_t} {\bf x}_{i_t}  \cdot {\bf x}_{j_t}  \\
  {\rm subject \:to} \\
  \sum\limits_{i_t} {\alpha} _{i_t} l_{i_t}  = 0 \\ 
 {\alpha} _{i_t}  \ge 0, 
 \end{array}
\end{equation}
in which 
\begin{equation}
\label{w}
{\bf w} = \sum\limits_{i_t} {\alpha} _{i_t} l_{i_t} {\bf x}_{i_t }. 
\end{equation}
Those ${\bf x}_{i_t}$ with ${\alpha} _{i_t} > 0$ are called support vectors. By substituting \eqref{w} into \eqref{hyper}, the solution of separating hyperplane is:
\begin{equation}
\label{fihy}
{\mathop{\rm f}\nolimits} ({\bf x}) = \sum\limits_{i_t} {\alpha} _{i_t} l_{i_t} {\bf x}_{i_t }\cdot  {\bf x}  + b.
\end{equation}

The brilliance of the \eqref{fihy} is that it just relies on the inner product between training points and testing point. It allows SVM to be easily generalized to nonlinear SVM. If ${\mathop{\rm f}\nolimits}({\bf x})$ is not a linear function about the data, the nonlinear SVM can be obtained by introducing a kernel function : 
\begin{equation}
\label{kernell}
{\mathop{\rm k}\nolimits} ({\bf x}_{i_t} ,{\bf x} ) = \varphi ({\bf x}_{i_t} )\cdot  \varphi ({\bf x} ) 
\end{equation}
to implicitly map the original data into a higher dimensional feature space ${\boldsymbol F}$, where $\varphi$ is the mapping from original space to feature space. In ${\boldsymbol F}$, $ \varphi ({\bf x}_{i_t} )$ are linearly separable. The separating hyperplane in feature space ${\boldsymbol F}$ is easily generalized into the following form:
\begin{equation}
\label{ghyper}
{\mathop{\rm f}\nolimits} ({\bf x}) = \sum\limits_{i_t} {\alpha} _{i_t} l_{i_t} \varphi ({\bf x}_{i_t} ) \cdot\varphi ({\bf x})  + b = \sum\limits_{i_t} {\alpha} _{i_t} l_{i_t} {\mathop{\rm k}\nolimits} ({\bf x}_{i_t} ,{\bf x})  + b.
\end{equation}

By introducing the kernel function, the mapping $\varphi$ need not be explicitly known which reduces much of the computational complexity. For much more details, refer to ~\cite{burges1998tutorial}~\cite{smola2004tutorial}.

A function is a valid kernel if there exists a mapping $\varphi$ satisfying \eqref{kernell}. Mercer's condition ~\cite{cristianini2000introduction} gives us the condition  about what kind of functions are valid kernels. Actually,  some common used kernels  are as follows:  polynomial kernels
\begin{equation}
\label{poly}
{\mathop{\rm k}\nolimits} ({\bf x}_i ,{\bf x}_j ) = ({\bf x}_i  \cdot {\bf x}_j  + 1)^d, 
\end{equation}
radial basis kernels (RBF)
\begin{equation}
{\mathop{\rm k}\nolimits} ({\bf x}_i ,{\bf x}_j ) = \exp ( - \gamma \left\| {{\bf x}_i  - {\bf x}_j } \right\|^2 ),
\end{equation}
and neural network type kernels
\begin{equation}
{\mathop{\rm k}\nolimits} ({\bf x}_i ,{\bf x}_j ) = \tanh (({\bf x}_i  \cdot {\bf x}_j ) + b),
\end{equation}
in which the heavy-tailed RBF kernel is in the form of 
\begin{equation}
{\mathop{\rm k}\nolimits} ({\bf x}_i ,{\bf x}_j ) = \exp ( - \gamma \left\| {{\bf x}_i^a  - {\bf x}_j^a } \right\|^b ),
\end{equation}
and Gaussian RBF kernel is 
\begin{equation}
\label{rbf}
{\mathop{\rm k}\nolimits} ({\bf x}_i ,{\bf x}_j ) = \exp \left( { - \frac{{\left\| {{\bf x}_i  - {\bf x}_j } \right\|^2 }}{{2\sigma ^2 }}} \right).
\end{equation}

\section{Dimensionality Reduction}
\label{dr}

Dimensionality reduction is a very effective tool in machine learning field.

In the rest of this paper, assuming the original dimensionality data are a set of $M$ samples ${\bf x}_i  \in {\bf R}^N ,i = 1,2, \cdots M $, the reduced dimensionality samples of ${\bf x}_i$ are ${\bf y}_i  \in {\bf R}^K ,i = 1,2, \cdots M $, where $K <  < N$.  $x_{ij}$ and $y_{ij}$ are componentwise elements in ${\bf x}_i$ and ${\bf y}_i$, respectively.

\subsection{Principal Component Analysis}
\label{secpca}

PCA ~\cite{jolliffe2002principal} is the best-known linear dimensionality reduction method. PCA aims to find a subspace $\Omega $ which can maximally retain the variance of the original dataset. The basis of $\Omega $ is obtained by eigen-decomposition of covariance matrix. The procedure can be summarized into the following four steps.
\begin{enumerate}
\label{pcap}
\item Compute the covariance matrix of ${\bf x}_i$
\begin{equation}
\label{covariance}
{\bf C} = \frac{1}{M}\sum\limits_{i = 1}^M {({\bf x}_i  - {\bf u})} ({\bf x}_i  - {\bf u})^T 
\end{equation}
where ${\bf u} = \frac{1}{M}\sum\limits_{i = 1}^M {{\bf x}_i }$ is the mean of the given samples, $T$ means transpose. 
\item Calculate eigenvalues $\lambda _1  \ge \lambda _2  \ge  \cdots  \ge \lambda _N $ and the corresponding eigenvectors ${\bf v}_1 ,{\bf v}_2 , \cdots ,{\bf v}_N $ of the covariance matrix $\bf C$. 
\item The basis of $\Omega $ is ${\bf v}_1 ,{\bf v}_2 , \cdots ,{\bf v}_K $.
\item Dimensionality reduction by 
\begin{equation}
y_{ij}  = ({\bf x}_i  - {\bf u}) \cdot {\bf v}_j.
\end{equation}
\end{enumerate}
The value of $K$ is determined by the criteria 
\begin{equation}
\frac{{\sum\limits_{i = 1}^K {\lambda _i } }}{{\sum\limits_{i = 1}^N {\lambda _i } }} > {\rm threshold}.
\end{equation}
Usually ${\rm threshold = 0}{\rm .95 \: or \: 0}{\rm .90}$.

\subsection{Kernel Principal Component Analysis} 
\label{seckpca}

PCA works well for the high dimensionality data with linear variability, but always fails when nonlinear nature exists. KPCA ~\cite{schölkopf1998nonlinear} is, on the other hand, designed to extract the nonlinear structure of the original data. It uses the  kernel function ${\mathop{\rm k}\nolimits}$ (same as SVM) to implicitly map the original data into a feature space ${\boldsymbol F}$, where $\varphi$ is the mapping from original space to feature space. In ${\boldsymbol F}$, PCA algorithm can work well. 

If ${\mathop{\rm k}\nolimits}$ is valid kernel function, the matrix 
\begin{equation}
{\bf K} = ({\mathop{\rm k}\nolimits} ({\bf x}_i ,{\bf x}_j ))_{i,j = 1}^M 
\end{equation}
 must be positive semi-definite. The matrix ${\bf K}$ is the so-called kernel matrix.

Assuming the mean of feature space data $\varphi ({\bf x}_i ),i = 1,2, \cdots M$ is zero, i.e.,
\begin{equation}
\frac{1}{M}\sum\limits_{i = 1}^M {\varphi ({\bf x}_i )}  = 0. 
\end{equation}
The covariance matrix in ${\boldsymbol F}$ is 
\begin{equation}
{\bf C}_F  = \frac{1}{M}\sum\limits_{i = 1}^M {\varphi ({\bf x}_i )\varphi ({\bf x}_i )^T }.
\end{equation}
In order to apply PCA in ${\boldsymbol F}$, the eigenvectors ${\bf v}_i^F $of ${\bf C}_F$ are needed. As we know that the mapping $\varphi$ is not explicitly known, thus the eigenvectors of ${\bf C}_F$ can not be as easily derived as PCA. However, the eigenvectors ${\bf v}_i^F $of ${\bf C}_F$ must lie in the span ~\cite{schölkopf1998nonlinear} of $\varphi ({\bf x}_i ),i = 1,2, \cdots M$, i.e.,
\begin{equation}
\label{eigenvec_feature}
{\bf v}_i^F  = \sum\limits_{j = 1}^M {\alpha _{ij} } \varphi ({\bf x}_j ).
\end{equation}
It has been proved that ${\boldsymbol \alpha }_i ,i = 1,2, \cdots ,M$ are eigenvectors of kernel matrix ${\bf K}$ ~\cite{schölkopf1998nonlinear}. In which $\alpha _{ij}$ are componentwise elements of ${\boldsymbol \alpha }_i$.

Then the procedure of KPCA can be summarized into the following six steps:
\begin{enumerate}
\label{kpcap}
\item Choose a kernel function ${\mathop{\rm k}\nolimits} $.
\item Compute kernel matrix 
\begin{equation} 
\label{kernel}
{\bf K}_{ij}  = {\mathop{\rm k}\nolimits} ({\bf x}_i ,{\bf x}_j ).
\end{equation}
\item The eigenvalues $\lambda _1^{\bf K} \ge \lambda _2^{\bf K} \ge \cdots \ge \lambda _M^{\bf K} $ and the corresponding eigenvectors ${\boldsymbol \alpha }_1 ,{\boldsymbol \alpha }_2 , \cdots ,{\boldsymbol \alpha }_M $ are obtained by diagonalizing  ${\bf K}$.
\item Normalizing ${\bf v}_j^F $ by ~\cite{schölkopf1998nonlinear}
\begin{equation}
1 = \lambda _j^{\bf K} ({\boldsymbol \alpha }_j  \cdot {\boldsymbol \alpha }_j ).
\end{equation}
\item The normalized eigenvectors ${\bf v}_j^F,j = 1,2, \cdots ,K$ constitute the basis of a subspace in ${\boldsymbol F}$.
\item The projection of a training point ${\bf x}_i$ on ${\bf v}_j^F, j = 1,2, \cdots ,K$ is computed by
\begin {equation}
y_{ij}  = ({\bf v}_j^F ,{\bf x}_i) = \sum\limits_{n = 1}^M {\alpha _{jn} {\mathop{\rm k}\nolimits} ({\bf x}_n ,{\bf x}_i)}.
\end{equation}
\end{enumerate}
The idea of kernel in KPCA is exactly the same with kernels in SVM. All of  kernel functions in SVM can be employed in KPCA, too.

So far the mean of $\varphi ({\bf x}_i ),i = 1,2, \cdots M$ has been assumed to be zero. In fact, the zero mean data in the feature space are
\begin{equation}
\varphi ({\bf x}_i )-\frac{1}{M}\sum\limits_{i = 1}^M {\varphi ({\bf x}_i )}. 
\end{equation}
The kernel matrix for this centering or zero mean data can be derived by ~\cite{schölkopf1998nonlinear}
\begin{equation}
{\bf \tilde K} = {\bf K} - 1_M {\bf K} - {\bf K}1_M  + 1_M {\bf K}1_M 
\end{equation}
in which $(1_M )_{ij} : = 1/M$.

\subsection{Maximum Variance Unfolding}
\label{secmvu}

MVU ~\cite{weinberger2006unsupervised} approach will be applied in our experiments among all the manifold learning methods. Resorting to the help of optimization toolbox, MVU can learn the inner product matrix of ${\bf y}_i$ automatically by maximizing their variance subject to the constraints that ${\bf y}_i$ are centered and local distances of ${\bf y}_i$ are equal to the local distances of ${\bf x}_i$. Here the local distances represent the distances between ${\bf y}_i$ (${\bf x}_i$) and its $k$ nearest neighbors, in which $k$ is a parameter.

The intuitive explanation of this approach is that when an object such as string is unfolded optimally, the Euclidean distances between its two ends must be maximized. Thus the optimization objective function can be written as 
\begin{equation}
{\rm maximize}\sum\nolimits_{ij} {\left\| {{\bf y}_i  - {\bf y}_j } \right\|} ^2,
\end{equation}
subject to the constraints, 
\begin{equation}
\begin{array}{c}
  \sum\nolimits_i {{\bf y}_i }  = 0 \\ 
  \left\| {{\bf y}_i  - {\bf y}_j } \right\|^2  = \left\| {{\bf x}_i  - {\bf x}_j } \right\|^2 
  \:{{\rm when }} \: \eta _{ij}  = 1
 \end{array} 
 \end{equation}
in which $\eta _{ij}  = 1$ means ${\bf x}_i$ and ${\bf x}_j$ are $k$ nearest neighbors otherwise $\eta _{ij}  = 0$.

Apply inner product matrix 
\begin{equation}
{\bf I} = ({\bf y}_i  \cdot {\bf y}_j )_{i,j = 1}^M 
\end{equation} 
of ${\bf y}_i$ to the above optimization can make the model simpler. 
The procedure of MVU can be summarized as follows:
\begin{enumerate}
\item Optimization step: because ${\bf I}$ is an inner product matrix, it must be positive semi-definite. Thus the above optimization can be reformulated into the following form ~\cite{weinberger2006unsupervised}
\begin{equation}
\label{op}
\begin{array}{l}
 {\rm maximize}\:{\rm trace}{\rm  }({\bf I} {\rm )} \\
 {\rm subject \: to}\\
    {\bf I}  \succ  0  \\ 
 \sum\nolimits_{ij} {{\bf I}_{ij} }  = 0  \\ 
 {\bf I}_{ii}  - 2{\bf I}_{ij}  + {\bf I}_{jj}  = D_{ij},{\rm  when } \:\eta _{ij}  = 1  \\ 
\end{array}
\end{equation}
where $D_{ij}= \left\| {{\bf x}_i  - {\bf x}_j } \right\|^2$, and ${\bf I}  \succ  0$ represents ${\bf I}$ is positive semi-definite.
\item The eigenvalues $\lambda _1^{\bf y}  \ge \lambda _2^{\bf y}  \ge  \cdots  \ge \lambda _M^{\bf y} $ and the corresponding eigenvectors ${\bf v}_1^{\bf y} ,{\bf v}_2^{\bf y} , \cdots ,{\bf v}_M^{\bf y}$ are obtained by diagonalizing  ${\bf I}$.
\item Dimensionality reduction by
\begin{equation}
y_{ij}  = \sqrt {\lambda _j^{\bf y} } v_{ij}^{\bf y} 
\end{equation}
in which $v_{ji}^{\bf y}$ are  componentwise elements of ${\bf v}_j^{\bf y}$.
\end{enumerate}

Landmark-MVU (LMVU) ~\cite{weinberger2005nonlinear} is a modified version of MVU which aims at solving larger scale problems than MVU. It works by using the inner product matrix ${\bf A}$ of randomly chosen landmarks from ${\bf x}_i$ to approximate the full matrix ${\bf I}$, in which the size of ${\bf A}$ is much smaller than ${\bf I}$. 

Assuming the number of landmarks is $m$ which are ${\bf a}_1,{\bf a}_2,\cdots,{\bf a}_m$, respectively. Let ${\bf Q}$ ~\cite{weinberger2005nonlinear} denote a linear transformation between landmarks and original dimensional data ${\bf x}_i  \in {\bf R}^N ,i = 1,2, \cdots M $, accordingly,
\begin{equation}
\label{transform}
\left( \begin{array}{c}
 {\bf x}_1  \\ 
 {\bf x}_2  \\ 
  \vdots  \\ 
 {\bf x}_M  \\ 
 \end{array} \right) \approx {\bf Q} \cdot \left( \begin{array}{c}
 {\bf a}_1  \\ 
 {\bf a}_2  \\ 
  \vdots  \\ 
 {\bf a}_m  \\ 
 \end{array} \right)
\end{equation}
in which
\begin{equation}
{\bf x}_i  \approx \sum\limits_j {{\bf Q}_{ij} {\bf a}_j } .
\end{equation}

Assuming the reduced dimensionality landmarks of ${\bf a}_1,{\bf a}_2,\cdots,{\bf a}_m$ are ${\bf \tilde y}_1 ,{\bf \tilde y}_2 , \cdots ,{\bf \tilde y}_m $ , and the reduced dimensionality samples of ${\bf x}_1,{\bf x}_2,\cdots,{\bf x}_M$ are ${\bf y}_1, {\bf y}_2,\cdots,{\bf y}_M $, then the linear transformation between ${\bf y}_1, {\bf y}_2,\cdots,{\bf y}_M $ and  ${\bf \tilde y}_1 ,{\bf \tilde y}_2 , \cdots ,{\bf \tilde y}_m $  is $\bf Q$ as well ~\cite{weinberger2005nonlinear}, consequently,
\begin{equation}
\left( \begin{array}{c}
 {\bf y}_1  \\ 
 {\bf y}_2  \\ 
  \vdots  \\ 
 {\bf y}_M  \\ 
 \end{array} \right) \approx {\bf Q} \cdot \left( \begin{array}{c}
 {\bf \tilde y}_1  \\ 
 {\bf \tilde y}_2  \\ 
  \vdots  \\ 
 {\bf \tilde y}_m  \\ 
 \end{array} \right).
\end{equation}

Matrix ${\bf A}$ is the inner-product matrix of ${\bf a}_1,{\bf a}_2,\cdots,{\bf a}_m$, 
\begin{equation}
{\bf A} = ({\bf \tilde y}_i  \cdot {\bf \tilde y}_j )_{i,j = 1}^m,
\end{equation} hence the relationship between ${\bf I}$ and ${\bf A}$ is
\begin{equation}
{\bf I} \approx {\bf QAQ}^T.
\end{equation}

The optimization of  \eqref{op} can be reformulated into the following form:
\begin{equation}
\label{lmvuop}
\begin{array}{l}
  {\rm maximize} \:{\rm trace} \:{\bf QAQ}^T  \\
   {\rm  subject\: to }\\
   {\bf A} \succ 0 \\ 
  \sum\nolimits_{ij} {({\bf QAQ}^T )_{ij} }  = 0 \\
  D_{ij}^{\bf y}  \le D_{ij},{\rm  when } \:\eta _{ij}  = 1
 \end{array}
\end{equation}
in which
\begin{equation}
D_{ij}= \left\| {{\bf x}_i  - {\bf x}_j } \right\|^2, 
\end{equation}
\begin{equation}
D_{ij}^{\bf y} =({\bf QAQ}^T )_{ii} - 2({\bf QAQ}^T )_{ij} + ({\bf QAQ}^T )_{jj},
\end{equation}
and ${\bf A} \succ  0$ represents ${\bf A}$ is positive semi-definite. This optimization model differs from \eqref{op} in that equality constraints for nearby distances are relaxed to inequality constraints in order to guarantee the feasibility of the simplified optimization model.

LMVU can increase the speed of programming but with the cost of decreasing accuracy. In  this paper's simulation, the LMVU will be applied.

\section{The Procedure of Using SVM and Dimensionality Reduction}
\label{ss}

As aforementioned, in radar and sensing signal processing, the control of DOF---or dimensionality---is the first step, called pre-processing. Both the linear and nonlinear methods will be investigated in this paper as pre-processing tools to extract the intrinsic dimensionality of the collected data. 


First, SVM will be employed for classification in cognitive radio. The classification power of SVM is tested by the testing set. The procedure of SVM for classification is summarized as follows.
\begin{enumerate}
\item The collected dataset ${\bf x}_i, i=1,2,\cdots,M$ will be divided into training sets and testing sets. The training samples and testing samples are ${\bf x}_i$ with indexes $i_t, t=1,2,\cdots,T$ and $i_s,s=1,2,\cdots,S$, respectively.
\item The labels $l_{i_t }$ and $l_{i_s }$ for ${\bf x}_{i_t }$ and ${\bf x}_{i_s }$ are extracted.
\item  Choose a kernel function from \eqref{poly} to \eqref{rbf}, and the corresponding parameter's values for the chosen kernel are designated.
\item The separating hyperplane in higher dimensional feature space ${\boldsymbol F}$ which is in the form of \eqref{ghyper} is trained by the training set ${\bf x}_{i_t }$.
\item The classification performance of the trained hyperplane  will be tested by the testing set ${\bf x}_{i_s }$.
\end{enumerate}
The above process will be repeated to gain averaged test errors.
 
\begin{figure}[!t]
\begin{center}
\scalebox{.30}{\includegraphics{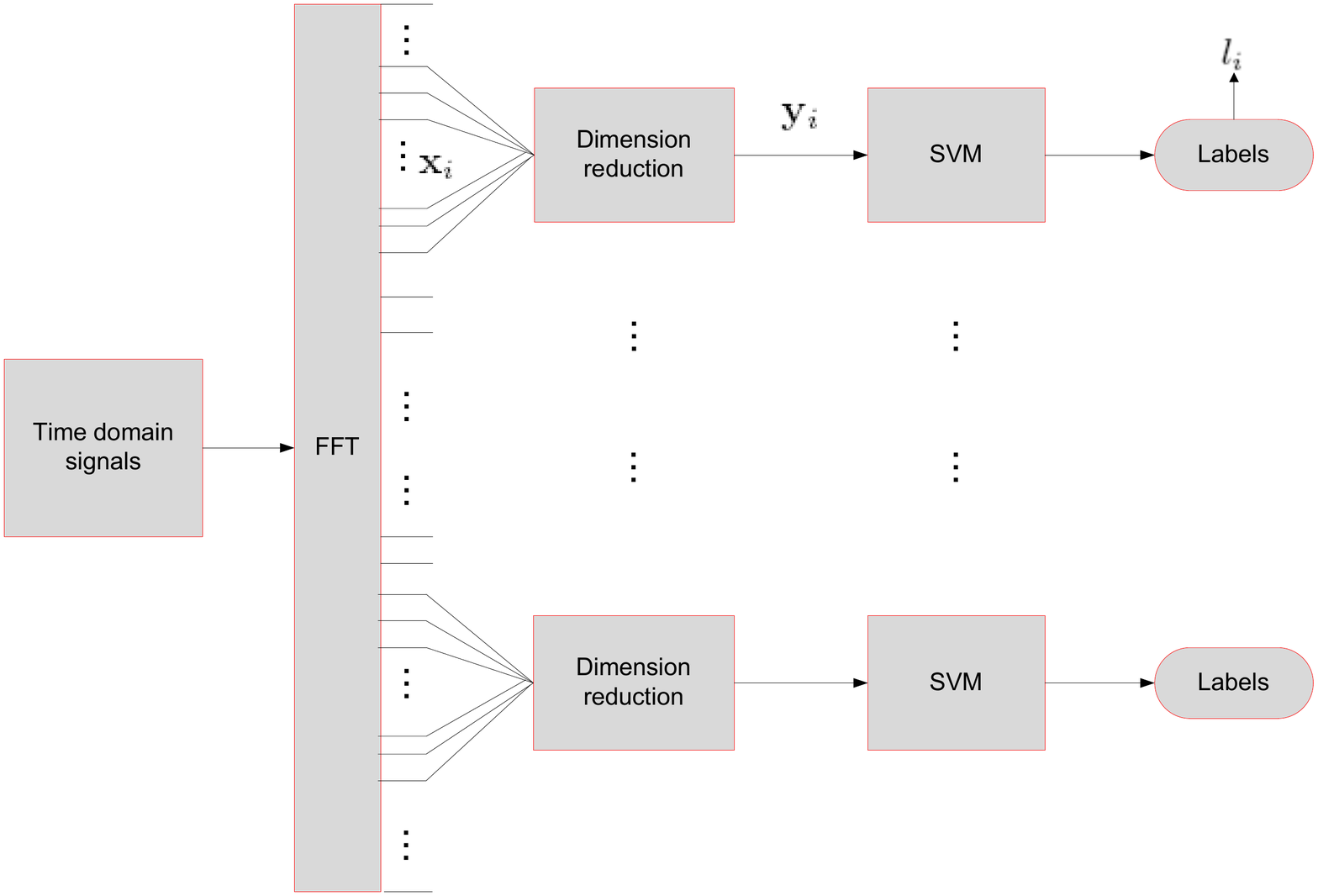}}
\end{center}
\vspace{-1ex}
\caption{The flow chart of SVM combined with dimensionality reduction for classification}
\vspace{-2ex}
\label{fig:schematic}
\end{figure}

Apart from applying SVM to classification, dimensionality reduction will be implemented before SVM  to get rid of the redundant information of the collected data.  The procedure of the proposed algorithm that is SVM combined with dimensionality reduction can be summarized as follows:
\begin{enumerate}
\item The collected dataset ${\bf x}_i, i=1,2,\cdots,M$ will be divided into training sets and testing sets. The training samples and testing samples are ${\bf x}_i$ with indexes $i_t, t=1,2,\cdots,T$ and $i_s,s=1,2,\cdots,S$, respectively.
\item The labels $l_{i_t }$ and $l_{i_s }$ for ${\bf x}_{i_t }$ and ${\bf x}_{i_s }$ are extracted.
\item Obtain reduced dimension data ${\bf y}_{i_t }$ and ${\bf y}_{i_s }$ by using of dimensionality reduction methods.
\item ${\bf y}_{i_t }$ and ${\bf y}_{i_s }$ are taken as the new training set and testing set.
\item  The labels $l_{i_t }$ and $l_{i_s }$ for ${\bf y}_{i_t }$ and ${\bf y}_{i_s }$ are kept unchanged with ${\bf x}_{i_t }$ and ${\bf x}_{i_s }$.
\item  Choose a kernel function from \eqref{poly} to \eqref{rbf}, and the corresponding parameter's values for the chosen kernel are designated.
\item The separating hyperplane in higher dimensional feature space ${\boldsymbol F}$ which is in the form of \eqref{ghyper} is trained by the new training set ${\bf y}_{i_t }$.
\item The classification performance of the trained hyperplane  will be tested by the new testing set ${\bf y}_{i_s }$.
\end{enumerate}
The above process will be repeated to gain averaged test errors. The flow chart of SVM combined with dimensionality reduction methods for classification is shown in Fig. \ref{fig:schematic}.

Dimensionality reduction methods of PCA, KPCA and MVU  are systematically studied in this paper which can meet the needs of all kinds of  data owning different structures.
Dimensionality reduction with PCA to derive ${\bf y}_{i_t }$ and ${\bf y}_{i_s }$ is implemented as follows.
\begin{enumerate}
\item The training set ${\bf x}_{i_t }$ is input to PCA procedure from step 1) to step 3) in section \ref{secpca} to obtain  the eigenvectors of ${\bf v}_1 ,{\bf v}_2 , \cdots ,{\bf v}_K $.
\item Dimensionality reduction by 
\begin{equation}
{\bf y}_{i_t }^T  = {({\bf x}_{i_t }  - {\bf u})}\cdot ({\bf v}_1,{\bf v}_2,\cdots,{\bf v}_K),
\end{equation}
\begin{equation}
{\bf y}_{i_s }^T  = {({\bf x}_{i_s } - {\bf u})} \cdot ({\bf v}_1,{\bf v}_2,\cdots,{\bf v}_K).
\end{equation}
\end{enumerate}

Dimensionality reduction with  KPCA to derive ${\bf y}_{i_t }$ and ${\bf y}_{i_s }$ is implemented as follows.
\begin{enumerate}
\item The training set ${\bf x}_{i_t }$ is input to KPCA procedure from step 1) to step 5) in section \ref{seckpca}  to obtain eigenvectors ${\bf v}_j^F,j = 1,2, \cdots ,K$, implicitly.
\item Dimensionality reduction by 
\begin{equation}
\begin{array}{ccc}
{\bf y}_{i_t }^T  &=& (({\bf v}_1^F, \cdots, {\bf v}_K^F)\cdot{\bf x}_{i_t }) \\
&=& (\sum\limits_{n = 1}^T {\alpha _{1n} {\mathop{\rm k}\nolimits} ({\bf x}_n ,{\bf x}_{i_t })},\\
 &&\cdots,\sum\limits_{n = 1}^T {\alpha _{Kn} {\mathop{\rm k}\nolimits} ({\bf x}_n ,{\bf x}_{i_t })}),
\end{array}
\end{equation}
\begin{equation}
\begin{array}{ccc}
{\bf y}_{i_s }^T  &=& (({\bf v}_1^F, \cdots, {\bf v}_K^F)\cdot{\bf x}_{i_s })\\
&=& (\sum\limits_{n = 1}^T {\alpha _{1n} {\mathop{\rm k}\nolimits} ({\bf x}_n ,{\bf x}_{i_s })},\\
&& \cdots,\sum\limits_{n = 1}^T {\alpha _{Kn} {\mathop{\rm k}\nolimits} ({\bf x}_n ,{\bf x}_{i_s })}).
\end{array}
\end{equation}
\end{enumerate}

Dimensionality reduction with LMVU to derive ${\bf y}_{i_t }$ and ${\bf y}_{i_s }$ is implemented as follows.
\begin{enumerate}
\item Both the training set and testing set ${\bf x}_{i_t }$ and ${\bf x}_{i_s }$ are input to MVU procedure from step 1) to step 2) in section  \ref{secmvu}  to obtain eigenvalues $\lambda _1^{\bf y}  \ge \lambda _2^{\bf y}  \ge  \cdots  \ge \lambda _M^{\bf y} $ and the corresponding eigenvectors ${\bf v}_1^{\bf y} ,{\bf v}_2^{\bf y} , \cdots ,{\bf v}_M^{\bf y}$
\item Dimensionality reduction by 
\begin{equation}
{\bf y}_{i_t }^T  = (\sqrt {\lambda _1^{\bf y} } v_{i_t1}^{\bf y},\cdots, \sqrt {\lambda _K^{\bf y} } v_{i_tK}^{\bf y}),
\end{equation}
\begin{equation}
{\bf y}_{i_s }^T  = (\sqrt {\lambda _1^{\bf y} } v_{i_s1}^{\bf y},\cdots, \sqrt {\lambda _K^{\bf y} } v_{i_sK}^{\bf y}).
\end{equation}
\end{enumerate}
To use LMVU, it is necessary to substitute \eqref{op} by \eqref{lmvuop} in the MVU procedure.

\section{Wi-Fi Signal Measurement}
\label{wifi}


\begin{figure}[!t]
	\centering
	\includegraphics[width=2.5in]{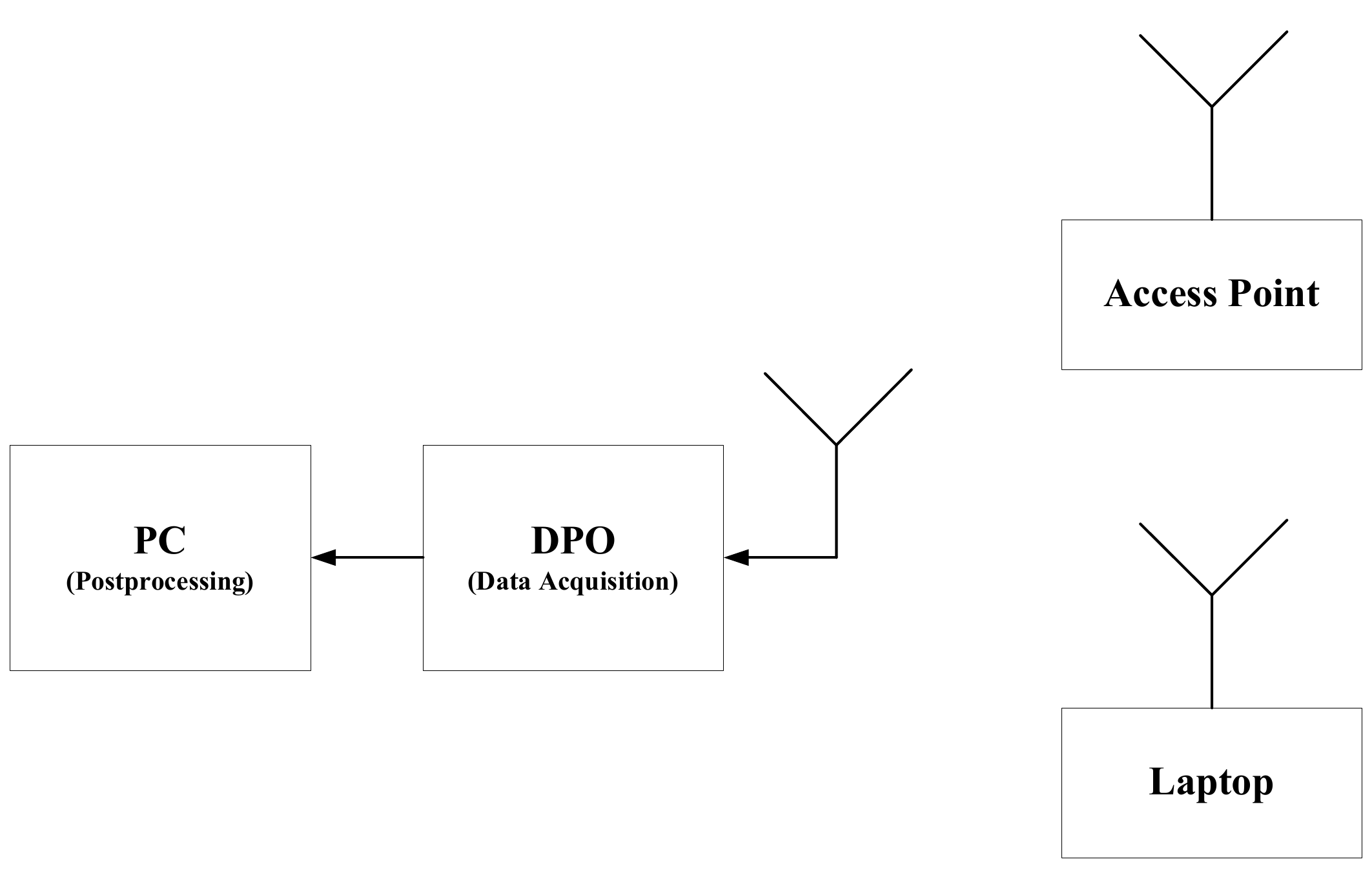}
	\caption{Setup of the measurement of Wi-Fi signals.}
	\label{fig_setup}
\end{figure}

\begin{figure}[!t]
	\centering
	\includegraphics[width=2.5in]{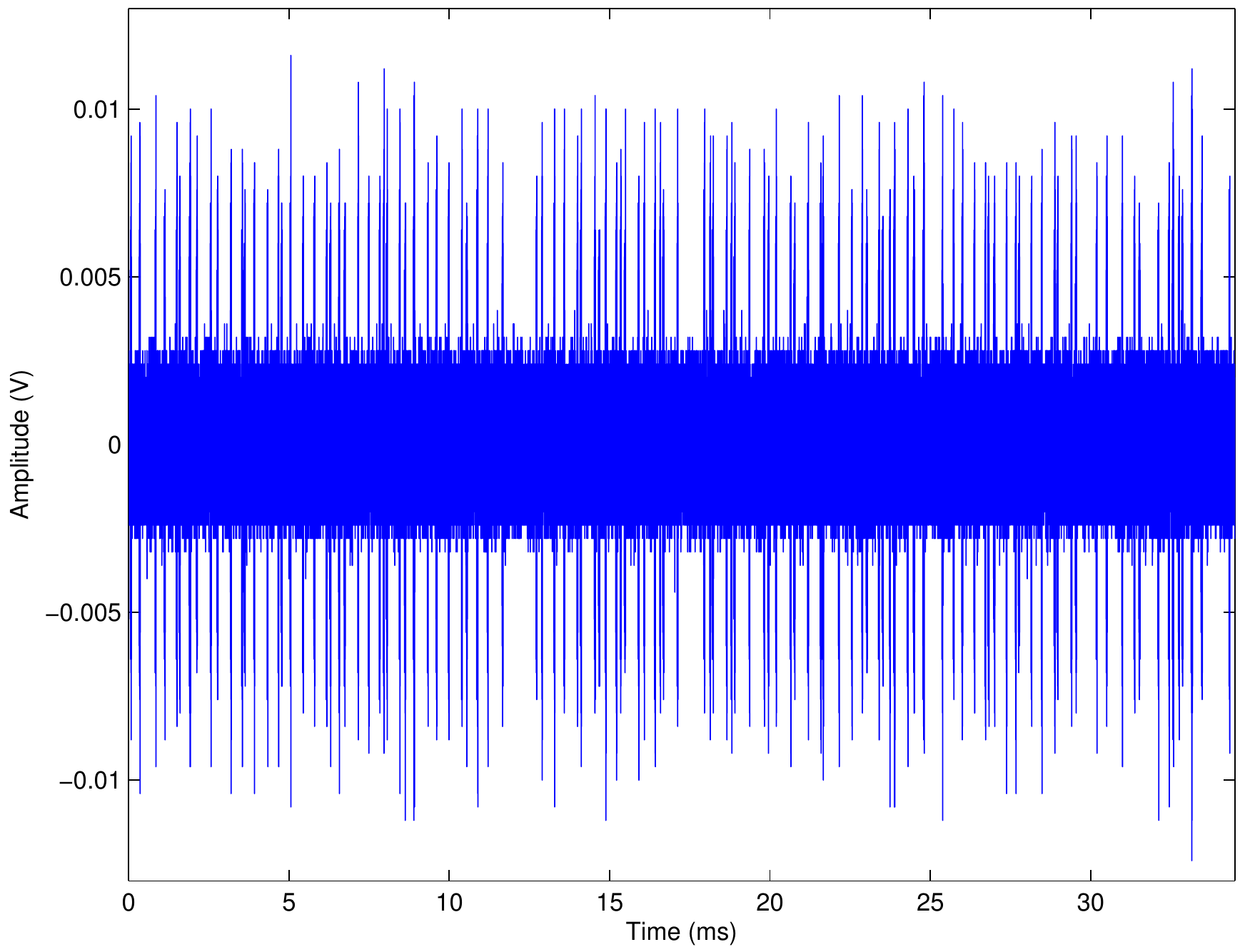}
	\caption{Recorded Wi-Fi signals in time-domain.}
	\label{fig_wifi}
\end{figure}




Wi-Fi time-domain signals have been measured and recorded using an advanced DPO whose model is Tektronix DPO72004~\cite{Chen2010hohmm}.
The DPO supports a maximum bandwidth of 20 GHz and a maximum sampling rate of 50 GS/s. It is capable to record up to 250 M samples per channel. In the measurement, a laptop accesses the Internet through a wireless Wi-Fi router, as shown in Fig.~\ref{fig_setup}. An antenna with a frequency range of 800 MHz to 2500 MHz is placed near the laptop and connected to the DPO. The sampling rate of the DPO is set to 6.25 GS/s. Recorded time-domain Wi-Fi signals are shown in Fig.~\ref{fig_wifi}. The duration of the recorded Wi-Fi signals is 40 ms.


The recorded 40-ms Wi-Fi signals are divided into 8000 slots, with each slot lasting 5 $\mu s$.  The time-domain Wi-Fi signals within the first 1 $\mu s$ of every slot are then transformed into frequency domain using fast Fourier transform (FFT). 
The spectral states of the measured Wi-Fi signal at each time slot present two possibilities. One is that current spectrum is occupied (state is busy $l{}_i=1$) at this time slot or current spectrum is  unoccupied (state is idle $l{}_i=0$). 
In this paper, the frequency band of 2.411 - 2.433 GHz is considered. The resolution in frequency domain is 1 MHz. Thus, for each slot, 23 points in frequency domain can be obtained. The total obtained data are shown in Fig.~\ref{fig:experiment_data}.

\section{Experimental Validation}
\label{ev}

\begin{figure}[!t]
\begin{center}
\scalebox{.35}{\includegraphics{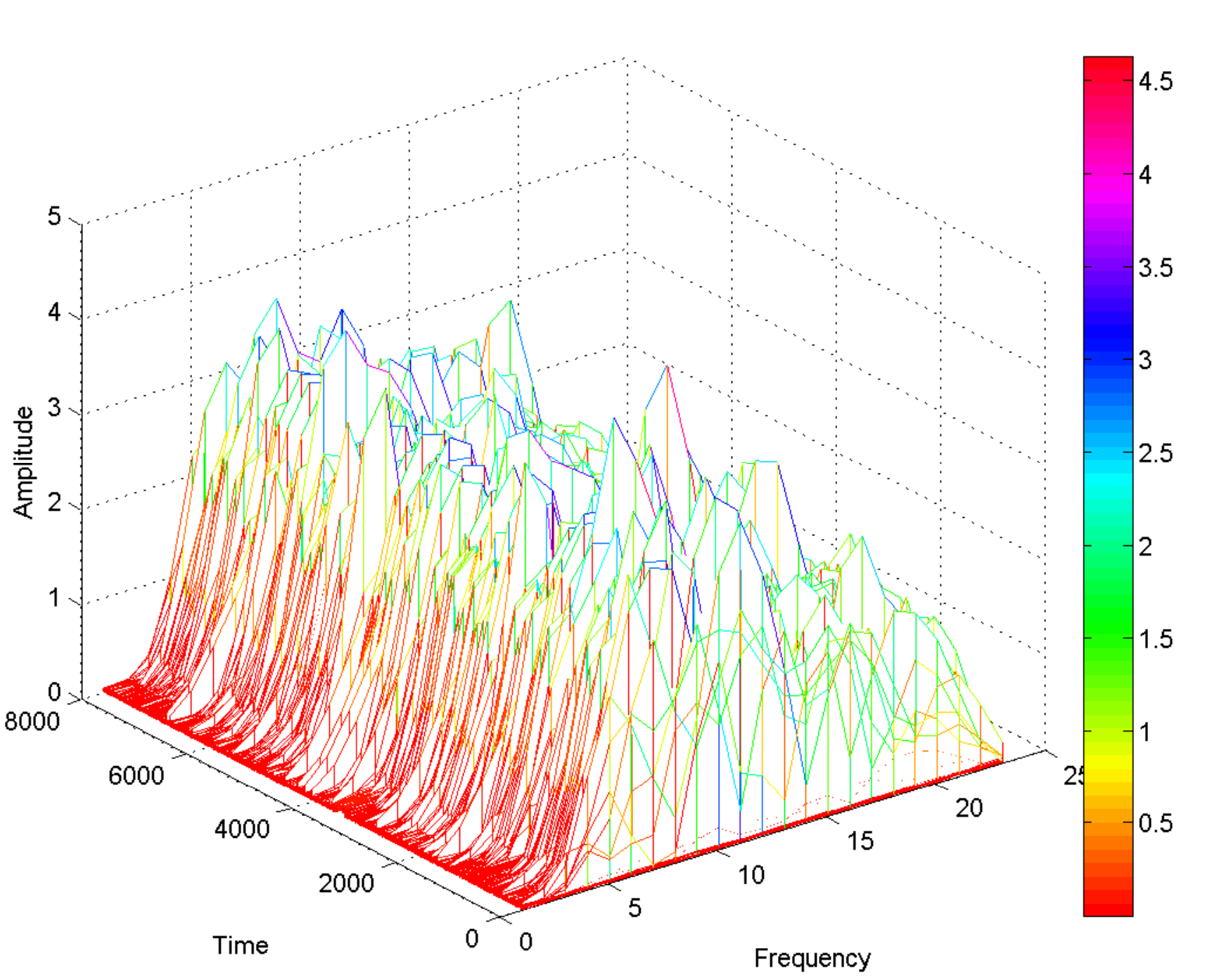}}
\end{center}
\vspace{-1ex}
\caption{The experimental data.}
\vspace{-2ex}
\label{fig:experiment_data}
\end{figure}

The spectral states of all time slots for measured Wi-Fi signal can be divided into two classes (busy $l{}_i=1$ or idle $l{}_i=0$). The powerful classification technique in machine learning, SVM, can be employed to classify the data at each time slot.
The processed Wi-Fi data is shown in Fig. \ref{fig:experiment_data}. 
The data used below are from the $1101^{{\rm th}}$ time slot to the $8000^{{\rm th}}$ time slot. 

In the next experiment, amplitude values in frequency domain of $i^{{\rm th}}$  usable time plot are taken as ${\bf x}_i$. 
The experiment data is taken by
\begin{equation}
\label{data}
\left( \begin{array}{c}
 {\bf x}_1  \\ 
 {\bf x}_2  \\ 
  \vdots  \\ 
 {\bf x}_{{\rm tot}}  \\ 
 \end{array} \right) = \left[ \begin{array}{c}
 x_{1,12 - m} ,...,x_{1,12} ,...,x_{1,12 + n}  \\ 
 x_{2,12 - m} ,...,x_{2,12} ,...,x_{2,12 + n}  \\ 
  \vdots  \\ 
 x_{{\rm tot},12 - m} ,...,x_{{\rm tot},12} ,...,x_{{\rm tot},12 + n}  \\ 
 \end{array} \right]
\end{equation}
where $x_{ij}$ represents amplitude value on the $j^{{\rm th}}$ frequency point of the $i^{{\rm th}}$ time slot. 
The dimension of ${\bf x}_i$ is $N=n+m+1$, in which $0 \leq n-m \leq 1$.

\begin{figure}[!t]
\begin{center}
\scalebox{.35}{\includegraphics{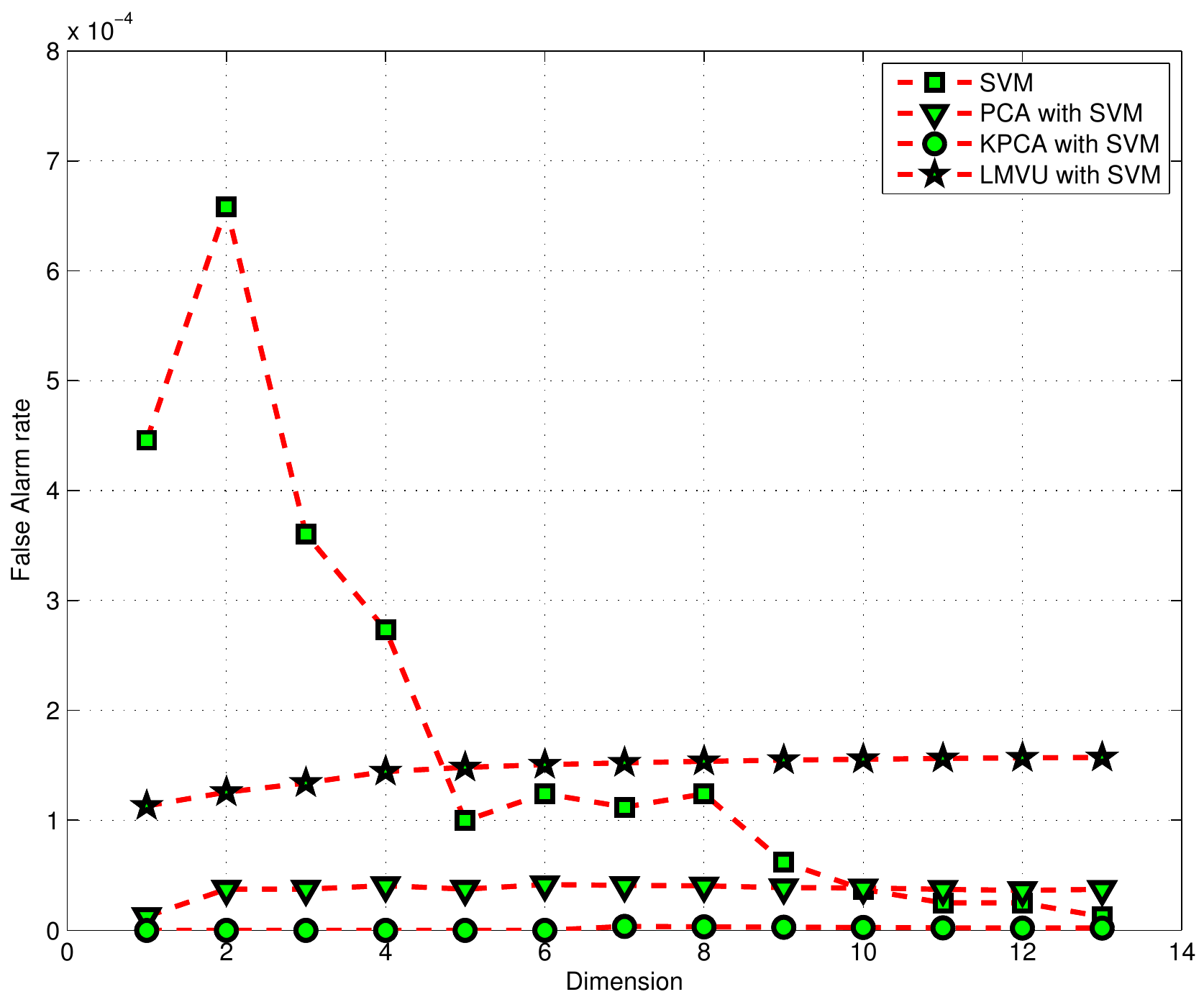}}
\end{center}
\vspace{-1ex}
\caption{False alarm rate.}
\vspace{-2ex}
\label{fig:false_alarm}
\end{figure}

\begin{figure}[!t]
\begin{center}
\scalebox{.35}{\includegraphics{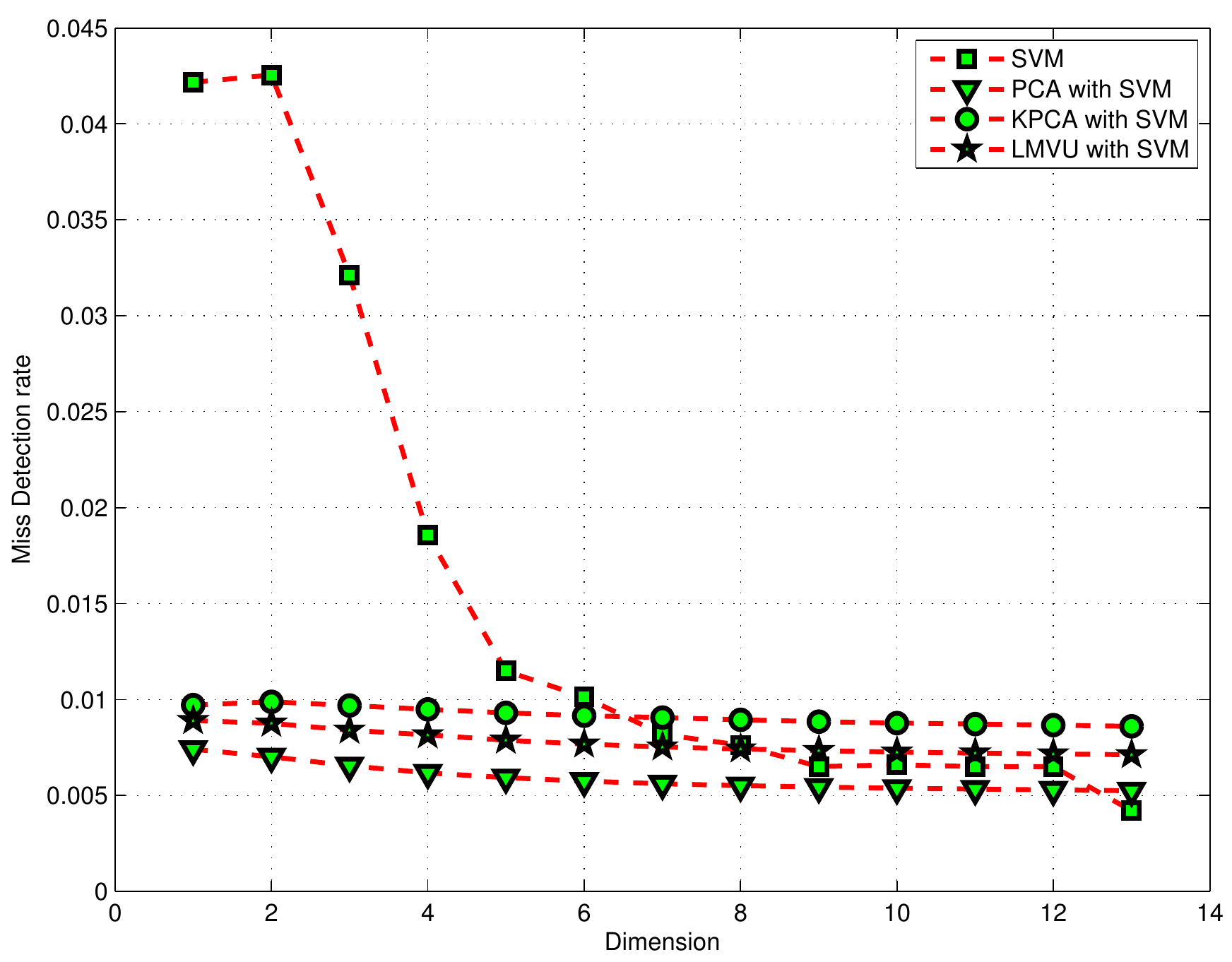}}
\end{center}
\vspace{-1ex}
\caption{Miss detection rate.}
\vspace{-2ex}
\label{fig:miss_detection}
\end{figure}

\begin{figure}[!t]
\begin{center}
\scalebox{.35}{\includegraphics{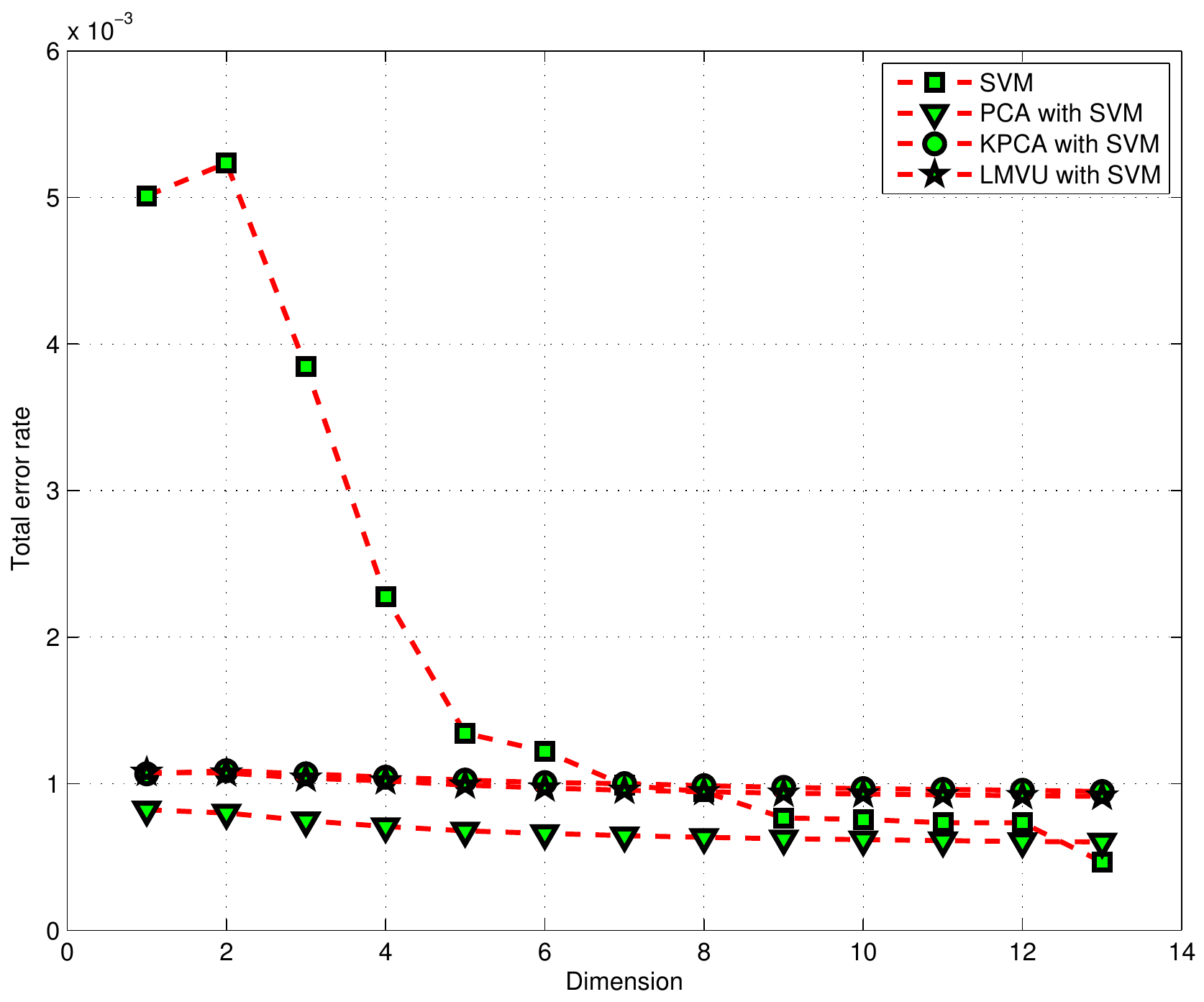}}
\end{center}
\vspace{-1ex}
\caption{Total error rate.}
\vspace{-2ex}
\label{fig:total_error}
\end{figure}

First, the true state $l_i$ of ${\bf x}_i$ is obtained.
The false alarm rate, miss detection rate and total error rate of experiments are shown in Fig. \ref{fig:false_alarm}, Fig. \ref{fig:miss_detection} and Fig. \ref{fig:total_error}, respectively, in which total error rates are the total number of miss detection and false alarm samples divided by the total number of testing set.  The results shown are the corresponding averaged values of 50 experiments. In each experiment, the number of training set is 200 and the number of testing set is 1800.


In these results, the steps of the ``SVM'' method for the $j^{{\rm th}}$ experiment are: 
\begin{enumerate}
\item Training set and testing set are chosen denoted by ${\bf x}_{i_t }$ and ${\bf x}_{i_s }$.
\item ${\bf x}_{i_t }$ and ${\bf x}_{i_s }$ are taken by \eqref{data} with dimensionality $N = 1,2,3,...,13$, in which $0 \leq n-m \leq 1$.
\item For each dimension $N$, ${\bf x}_{i_t }$ and ${\bf x}_{i_s }$ are fed to SVM algorithm. 
\end{enumerate}
The above process repeats with $j = 1,2,...,50$, then the corresponding averaged values of these 50 experiments are derived for each dimension.

However, for the methods of ``PCA with SVM'', ``KPCA with SVM'' and ``LMVU with SVM'', their steps for the $j^{{\rm th}}$ experiment are:
\begin{enumerate}
\item Training set and testing set are chosen denoted by ${\bf x}_{i_t }$ and ${\bf x}_{i_s }$.
\item ${\bf x}_{i_t }$ and ${\bf x}_{i_s }$ are taken by \eqref{data} with dimensionality $N=13$, in which $n=m=6$.
\item ${\bf y}_{i_t }$ and ${\bf y}_{i_s }$ are reduced dimensional samples of ${\bf x}_{i_t }$ and ${\bf x}_{i_s }$ with PCA, KPCA and LMVU methods, respectively.
\item The dimensions of ${\bf y}_{i_t }$ and ${\bf y}_{i_s }$ vary by $K = 1,2,3,...,13$ manually.
\item  For each dimension $K$, ${\bf y}_{i_t }$ and ${\bf y}_{i_s }$ are fed to SVM algorithm.
\end{enumerate}
The above process repeats with $j = 1,2,...,50$, then the corresponding averaged values of these 50 experiments are derived for each dimension. 

In fact, for LMVU approach, both the placements and the number of landmarks can influence its performance. The choice of landmarks for each experiment is as follows. For every experiment, the number of landmarks $m$ is equal to 20.  At the beginning of the LMVU process, ten groups of randomly chosen positions in ${\bf x}_i$ (including both training and testing sets) are obtained with 20 positions in each group, and then these positions are fixed. For each experiment, results with landmark positions assigned by each group are obtained. Given ten groups of landmarks, the group of which that can get minimal total error rate is taken as the landmarks for this experiment.

In this whole experiment, Gaussian RBF kernel with $2\sigma ^2  = 5.5^2$ is used for KPCA. The parameter $k=3$, in which $k$ is the number of nearest neighbors of ${\bf y}_i$ (${\bf x}_i$) (including both training and testing sets) for LMVU. The optimization toolbox ${\rm SeDuMi\: 1.1R3}$ ~\cite{sturm2006advanced} is applied to solve the optimization step in LMVU. The SVM toolbox SVM-KM ~\cite{SVM-KMToolbox} is used to train and test SVM processes. The kernels selected for SVM are heavy-tailed RBF kernels with parameters 
${\gamma} = 1,a=1,b=1$. These parameters keep unchanged for the whole experiment.

\begin{figure}[!t]
\begin{center}
\scalebox{.25}{\includegraphics{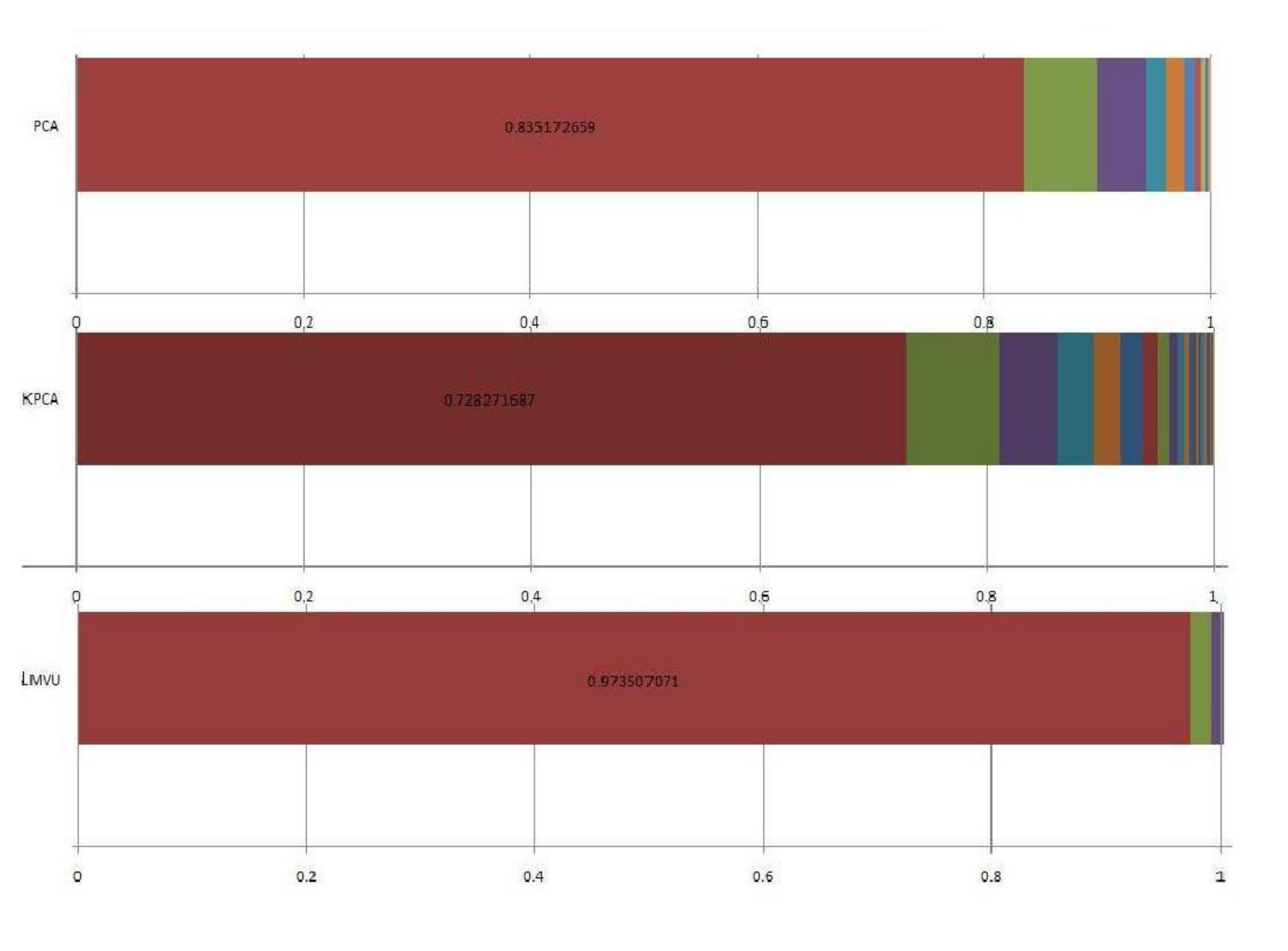}}
\end{center}
\vspace{-1ex}
\caption{Distributions of corresponding eigenvalues.}
\vspace{-2ex}
\label{fig:eigenvalue}
\end{figure}

Take data set of the first experiment as an example. The distributions of corresponding normalized (the summation of total eigenvalues equals one) eigenvalues of PCA, KPCA and LMVU methods are shown in Fig. \ref{fig:eigenvalue}. It can be seen that  the largest eigenvalues for the three methods  are  all dominant  ($84\%, 73\%,97\%$ for PCA, KPCA and LMVU, respectively). Consequently,  reduced one-dimensional data can even extract most of the useful information  from the original data. 

As can be seen from Fig. \ref{fig:false_alarm}, Fig. \ref{fig:miss_detection} and Fig. \ref{fig:total_error}, with the only use of SVM, the classification errors are very low. It means SVM successfully classifies the states of the Wi-Fi signal.  The results also verify the fact that more and more dimensions of the data are included (more information are included), the error rates should become smaller and smaller globally.
On the other hand, the results with ${\bf y}_i$ fed to SVM  can gain a higher accuracy at lower dimensionality than with ${\bf x}_i$ directly fed to SVM. 
By dimensionality reduction, most of the useful information including in  ${\bf x}_i$ can be extracted to the first $K= 1$ dimension of ${\bf y}_i$. Therefore, even if we increase the dimensions of the reduced data, the error rates do not embody obvious improvement. The error rates with only one feature of the proposed algorithm can match the error rates of 13 features of the original data.

%



\section{Conclusion}
\label{conc}

One fundamental open problem is to determine how and where machine learning algorithms are useful in a cognitive radio network.  The network dimensionality has received attention in information theory literature. One naturally wonders how network (signal) dimensionality affects the performance of system operation. Here we study, as an illustrative example, spectral states classification under this context.  Both linear (PCA) and nonlinear methods (KPCA and LMVU) are studied, by combining them with SVM.


Experimental results show that data with only one feature fed to SVM, the false alarm rate of method with dimensionality reduction is at worst $0.03068\%$ comparing with $0.06581\%$ of method without dimensionality reduction, and the miss detection rate is  $0.9883\%$ comparing with $4.255\%$. The error rates with only one feature of the methods with dimensionality reduction can nearly match the error rates of 13 features of the original data. 

The results of only appling SVM  verify the fact that more and more dimensions of the data are included, the error rates should become smaller and smaller globally since more information of the original data are considered. However, SVM combined with dimensionality reduction does not embody such property. This is because that the reduced dimension data with only one dimension already extracts most of the information of the original Wi-Fi signal. Therefore, even if we increase the dimensions of the reduced data, the error rates do not embody obvious improvement.

In this paper, SNR of the measured Wi-Fi signal is high which makes the error rates of the classification all very low.  In fact, dimensionality reduction  can not only  get rid of the redundant information but also have the effect of de-noising. When the collected data contains more noise, SVM combined with dimensionality reduction should embody more advantages. Besides, the original dimension of the spectral domain Wi-Fi signal is not high which also makes the advantage of dimensionality reduction not obvious.  In the future, dimensionality reduction used as pre-processing tool will be applied to more scenarios in cognitive radio such as low SNR (spectrum sensing) and very high original dimensions ( cognitive radio network) contexts. 

In the MVU approach, LMVU method is used in this paper which decreases the accuracy. On the other hand, the optimization toolbox ${\rm SeDuMi\: 1.1R3}$ is used to solve the optimization step which slows down the computation.  The dedicated optimization  algorithm for MVU should be proposed in the future. 

In the future, more machine learning algorithms will be, systematically, explored and applied to the cognitive radio network under different scenarios. 

\section*{Acknowledgment}
This work is funded by National Science Foundation
through two grants (ECCS-0901420 and ECCS-0821658), and
Office of Naval Research through two grants (N00010-10-1-
0810 and N00014-11-1-0006).




%


\bibliographystyle{ieeetr}
\bibliography{dsoref/machinelearning,dsoref/Manifoldlearning,dsoref/cr_prediction}

\end{document}